\documentclass[aps,pra,superscriptaddress,twocolumn,showpacs,longbibliography]{revtex4-2}
\usepackage[utf8]{inputenc}
\usepackage[english]{babel} 
\usepackage[T1]{fontenc}
\usepackage{graphicx}
\usepackage{xcolor,etoolbox}
\usepackage{dcolumn}
\usepackage{bm}
\usepackage{amsmath,amsthm,amssymb,mathrsfs,mathtools,amsfonts,braket}
\usepackage{verbatim}
\usepackage{ulem}
\usepackage{color}
\usepackage{xfrac}

\usepackage[colorlinks=true,
  linkcolor=olive,
  citecolor=blue,
  urlcolor =blue]{hyperref}

\usepackage{listings}
\definecolor{backcolor}{rgb}{0.95,0.95,0.92}

\begin{document}
\title{
Pairwise Liouvillian learning from randomized measurements:
practical aspects and guidelines for operating the protocol in large-scale experiments}
\author{William T. Lam}
\affiliation{Univ.~Grenoble Alpes, CNRS, LPMMC,  Grenoble, France}

\author{Manoj K. Joshi}
\affiliation{Science, Mathematics and Technology Cluster, Singapore University of Technology and Design, 8 Somapah Road, 487372, Singapore}
\affiliation{Institute for Quantum Optics and Quantum Information, Austrian Academy of Sciences, Innsbruck, 6020, Austria}

\author{Daniel Stilck França}
\affiliation{Department of Mathematical Sciences and Quantum for Life Center,
University of Copenhagen, Copenhagen, Denmark}

\author{Beno\^it Vermersch}
\affiliation{Univ.~Grenoble Alpes, CNRS, LPMMC,  Grenoble, France}
\affiliation{Quobly, Grenoble, France}

\begin{abstract}
We review and numerically study a protocol for 
Liouvillian learning based on randomized Pauli states and measurements. 
In particular, in the two-body, long-range interactions, and single-body noise setting, we describe the complete workflow to obtain the coefficients of the Liouvillian in an efficient and 
pairwise manner, meaning that the required classical memory is independent of the system size.
We also provide guidelines for choosing the 
parameters for data acquisition and postprocessing
that minimize the total reconstruction error. 
\end{abstract}

\maketitle

\hypersetup{linkcolor=red}

\section{Introduction}\label{sec:introduction}
Analog quantum simulators (AQS) 
hold the promise to simulate quantum many-body problems  
in regimes that are challenging for classical computers~\cite{trivedi_quantum_2024}.
In such systems, the quantum particles (atoms, ions, superconducting circuits, spins, etc) interact via a many-body Hamiltonian.
The unwanted couplings to the environment (noise) are captured, in the Markovian approximation, by Lindblad operators, which, along with the Hamiltonian, form the Liouvillian, a linear operator that fully specifies the evolution of the many-body system.

In the context of AQS, one would like to program the Hamiltonian part of the Liouvillian, 
in order to study a quantum simulation problem (superconductivity, magnetism, etc),
let the system evolve for a certain time $t$, and measure meaningful quantum observables
to produce a phase diagram. 
The difference between the target Liouvillian
and the one that is actually experimentally
realized is a key metric to assess the success of a quantum simulation 
experiment, in particular in the regime of quantum advantage~\cite{trivedi_quantum_2024, cai_stochastic_2024, Kashyap2025, Rao2025, kraft_bounded-error_2025}.

In recent years, a plethora of protocols for learning the
Liouvillian of an experimental quantum system
have been proposed~\cite{stilck_franca_efficient_2024, franca_learning_2025, zubida_optimal_2021, ivashkov_ansatz-free_2026, olsacher_hamiltonian_2025, huang_learning_2023, yu_robust_2023,  zhao_learning_2025, Ma2024, bakshi_structure_2024, hu_ansatz-free_2025, brahmachari_learning_2026} 
and implemented~\cite{kraft_bounded-error_2025, severin_learning_2026,  birke_demonstrating_2026, guo_hamiltonian_2025, franceschetto_hamiltonian_2025, berg_large-scale_2025, hangleiter_robustly_2024}.
Under some assumptions such as the locality of the Hamiltonian interactions,
such protocols are provably scalable, i.e. the Liouvillian learning can be achieved
with arbitrary accuracy using experimental and postprocessing resources that
scale polynomially with system size.
This makes Liouvillian learning the standard procedure to verify a quantum simulation 
experiment. 

While Liouvillian learning protocols are undoubtedly scalable, 
the actual experimental and postprocessing effort remain 
significant in a large-scale scenario.
In a recent work, in Ref.~\cite{kraft_bounded-error_2025}, 
Liouvillian learning and error-bounded quantum simulation 
was demonstrated in a trapped ion system of up to $N=51$ qubits.
In such article, we took advantage of a protocol~\cite{stilck_franca_efficient_2024}, 
based on randomized measurements~\cite{elben_randomized_2022}, 
where each parameter of the Liouvillian can be isolated, by picking a particular set of experimental configurations.
While Ref.~\cite{kraft_bounded-error_2025} focused on the demonstration of error-bounded quantum simulation in a specific experimental scenario, we would 
like here to show the general applicability of the protocol 
for large-scale experiments  and to provide concrete 
guidelines to use it with minimal resources. 
In particular, we explain an efficient data sampling and postprocessing workflow for reconstructing the Liouvillian pairwise, i.e., accessing the terms associated with each pair of qubits separately.
Our work provides in particular numerical studies 
of statistical errors and systematic errors 
that show how to practically 
choose all the key parameters 
of both the experimental and postprocessing steps.

This article is organized as follows. In Sec.~\ref{sec:method}, we review the details
of pairwise Liouvillian learning in the two-body interactions setting. 
In Sec.~\ref{sec:exp_friendly} we present the experimental task associated with the protocol that is based on Pauli randomized measurements, highlighting some 
differences with the method presented in Ref.~\cite{stilck_franca_efficient_2024}. Then, in Sec.~\ref{sec:poly_interp}, we explain how we perform the polynomial interpolation in practice.
In Sec.~\ref{sec:numerical_study}, we show numerical results on the influence of the settings on the reconstruction of the Liouvillian. 
Lastly, in Sec.~\ref{section:code}, we illustrate how to use our code to perform the learning protocol in an experiment.

\section{Theory of pairwise Liouvillian learning}
\label{sec:method}

In this section, we review the protocol~\cite{stilck_franca_efficient_2024},
and introduce a few modifications to it that will be important for what follows. 

\subsection{Liouvillian ansatz}\label{subsec:setting}
We consider a set of qubits $i=1,\dots,N$ whose density matrix $\rho$ evolves 
via a Lindblad equation
\begin{equation}
    \frac{d}{dt} \rho(t) = -\mathrm{i} \ [H, \rho(t)] +D[\rho(t)].
    \label{eq:eins}
\end{equation}
Our task consists in learning the Liouvillian $\{H,\mathcal{D}\}$ based on 
experimental measurements. 
Aligning with typical AQS experiments~\cite{monroe_programmable_2021, browaeys_many-body_2020, blatt_quantum_2012}, 
where engineered spin-spin interactions are considered to be up to two-body long-range models, 
we write $H$, in the Pauli basis, as 

\begin{equation}
\begin{split}
	H &= \sum_{i=1}^N \sum_{a}
	h_{i,a} \: \sigma_i^{(a)}
	 + \sum_{i < j} \sum_{a, b} h_{i,a,j,b} 
	 \ \sigma_i^{(a)} \sigma_j^{(b)},
	\label{H_model}
\end{split}
\end{equation}
with $a,b \in\{ x,y,z\}$ and $i,j$ the qubit indices. 
In total, we describe the Hamiltonian with
$3N+9N(N-1)/2$ real parameters. 

We consider single-body noise processes,
\textit{i.e.} processes that are only involving single Pauli matrices, 
that we write as
\begin{equation}
\begin{split}
	D[\rho] &= \sum_{i, j}  \sum_{a,b} d_{i,a,j,b} \ A(\sigma_i^{(a)}, \sigma_j^{(b)}, \rho), \\
	A(\sigma_i^{(a)}, \sigma_j^{(b)}, \rho) &= \sigma_i^{(a)} \rho \sigma_j^{(b)} - \frac{1}{2} \{ \sigma_j^{(b)} \sigma_i^{(a)}, \rho \} ,
	\label{D_model}
\end{split}
\end{equation}
with $a,b\in \{ x,y,z\}$.
This includes for instance models of local dephasing, amplitude damping, depolarization~\cite{Preskill_qinfo_chap3, nielsen_quantum_2010}, but also global dephasing that is relevant to experimental setups that use a global laser beam to control the qubits, such as trapped ions or Rydberg atoms  platforms~\cite{Bohnet_spin_dyn, kraft_bounded-error_2025,  carnio_generating_2016}. 

To give a physical interpretation of these parameters, 
we can construct and diagonalize~\cite{wolf_q_chanel} 
the $(3N \times 3N)$ positive semi-definite matrix
$\mathcal{D}_{(i,a),(j,b)}=d_{i,a,j,b}$.
This provides the jump operators (eigenvectors) that describe the physical dissipative processes
$
	L_{\nu}=\sum_{i} \sum_{a} l_{i,a} \sigma_i^{(a)},
$
with $\{ l_{i,a} \}$ their decomposition in the Pauli basis, 
and $\gamma_\nu$, the rates of the dissipative processes (eigenvalues).
Now we can build the \textit{diagonal} form of the Lindblad equation that describes the dissipation at best,
\begin{equation}
	D[\rho] = \sum_\nu \gamma_{\nu} \, A(L_\nu, L_\nu^\dagger, \rho).
	\label{D_diag}
\end{equation}

\subsection{Decoupling the system at time $t=0$}
Let us assume for the moment that we have access to the full continuous time sequence
$\braket{O(t)}=\mathrm{tr}(\rho(t) O)$ of certain observables $O$.
From Eq.~\eqref{eq:eins}, we obtain the equation at $t=0$
\begin{equation}
    \frac{d}{dt} \braket{O(t)}_{t=0}
	= \mathrm{tr}(-\mathrm{i}[H, \rho(0)]O) +\mathrm{tr}(D[\rho(0)]O).
    \label{lind}
\end{equation}
Writing this equation for several combinations of \mbox{$(O,\rho=\rho(0))$},
 we obtain
a linear system in terms of the coefficients of $H$ and $D$.
Therefore, solving this system allows us to realize Liouvillian learning. 

Let us now introduce a convenient parametrization of the operators
$O$ and $\rho$. We first select a given qubit pair $(i,j)$, $i<j$
and define the arrays of "configurations" $(\rho_c,O_c)$, $c=1,\dots,C$ 
of the type
 \begin{align}
 (\rho_c,O_c) &=
 (\rho_i,O_i)
 , (\rho_j,O_{j})
 ,(\rho_{ij},O_{ij}),
 \label{eq:configs}
 \end{align}
with $6$ possible values for single-body density matrices (up to reordering)
$\rho_k = \tau_k \otimes(\mathbf{1}/2)^{\otimes N-1}$, 
$\tau_k=(\mathbf{1}+\epsilon \sigma_k)/2$, 
$\epsilon=\pm 1$, and 
$\sigma_k = \sigma_k^{(x)}, \sigma_k^{(y)}, \sigma_k^{(z)}$, and 
$3$ possible values for each operator $O_k= \sigma_k^{(x)}, \sigma_k^{(y)}, \sigma_k^{(z)}$, for $k=i,j$; and similarly the two-body settings, 
$\rho_{ij}=\tau_i\otimes \tau_j \otimes(\mathbf{1}/2)^{\otimes N-2}$, $O_{ij} = \sigma_i \sigma_j$ can take $36$ and $9$ possible values. 

This gives in total at most $C_{\max}=360$ configurations to consider per qubit pair.
In what follows, and in contrast to Ref.~\cite{stilck_franca_efficient_2024}, we consider the general scenario $C\le C_{\max}$, having in mind experimental situations in which we prefer to measure only a finite subset of the possible configurations. 
Finally, we discuss in Sec.~\ref{sec:exp_friendly} how to effectively access the mixed states 
$\rho_i,\rho_j,\rho_{ij}$
by preparing random Pauli eigenstates. 

\subsection{Reduction to one and two qubits parameters}\label{reduction}
Now, let us plug the Hamiltonian in Eq.~\eqref{H_model} and the dissipative model in Eq.~\eqref{D_model} into Eq.~\eqref{lind}. 
Let us consider first a single-body setting $(\rho_i,O_i)$, and $\tau_i$ the associated Pauli state on $i$, and let us treat the two-body case after. 
Using the commutation rules of Pauli operators,
Liouvillian terms that are not acting on $i$ alone vanish~\cite{stilck_franca_efficient_2024}, 
\begin{equation}
\begin{split}
\text{tr}(-\mathrm{i}[\sigma_{k}^{(a)},\tau_{i}]O_{i}) &= 0 \text{, if } k \neq i, \\
	\text{tr}(A(\sigma_{k}^{(a)},\sigma_{k'}^{(b)}, \tau_{i}) O_{i}) 
	&= 0 \text{, if } k \neq k' \neq i, \\
\text{tr}(-\mathrm{i}[\sigma_{k}^{(a)}\sigma_{k'}^{(b)},\tau_{i}\otimes(\mathbf{1}/2)^{\otimes N-1}]O_{i}) &= 0, \forall k, k', 
\end{split}
\label{eq:vanish}
\end{equation}
such that we obtain (the derivative is still at $t=0$, but we omit the notation for the rest of the work), 
\begin{equation}
\label{eq:system1}
\begin{split}
    	\frac{d\braket{O_i}}{dt} 
	=
	\sum_{a} \mathrm{tr}& \left( -\mathrm{i}[\sigma_i^{(a)}, \tau_i] O_i \right) h_{i,a} \\
	+ \sum_{a,b} \text{tr}& \left( A(\sigma_i^{(a)},\sigma_i^{(b)}, \tau_i) O_i \right)  d_{i,a,i,b},
\end{split}
\end{equation}
which is a function of the $3$ single-qubit Hamiltonian terms
$\{h_{i,a}\}$ and the $9$ local dissipative terms $\{ d_{i,a,i,b} \}$, that are only acting on $i$. 

Similarly, with configurations of the type 
\mbox{$(\rho_{ij},O_{ij})$, $i \neq j$}, 
only the single-body Hamiltonian and dissipative terms acting on $i$ or $j$, and the two-body terms acting on the qubit-pair $(i,j)$ survive. We obtain

\begin{equation}
\begin{split}
\frac{d\braket{O_{ij}}}{dt} =
 \sum_{k=i,j} \sum_a &\mathrm{tr} \left( -\mathrm{i} [\sigma_k^{(a)}, \tau_i\otimes \tau_j] O_{ij} \right) h_{k,a} \\
+\sum_{a, b} &\mathrm{tr} \left( -\mathrm{i}[\sigma_i^{(a)} \sigma_j^{(b)}, \tau_i\otimes \tau_j]O_{ij} \right) h_{i,a,j,b} \\
+ \sum_{k,n=i,j}\sum_{a, b} &\mathrm{tr}\left( A(\sigma_k^{(a)}, \sigma_n^{(b)},
 \tau_i\otimes \tau_j)O_{ij} \right) d_{k,a,n,b},
  \label{eq:system2}
\end{split}
\end{equation}
which is a function of $6+9+4\times 9=51$ Hamiltonian and dissipative terms.

We note here that if we would have considered three-body or higher interaction terms in the Hamiltonian ansatz, we would still have obtained the same equations Eqs.~\eqref{eq:system1}-\eqref{eq:system2}, and so, still correctly learn the single- and two-body terms. To detect (and so to measure) higher-body Hamiltonian terms, one has to include higher body configurations $(\rho_c, O_c)$, see Ref.~\cite{stilck_franca_efficient_2024}. 
This is a key point of this learning method, as one can effectively learn the $l$-body Hamiltonian terms without it being affected by the ($l+1$)-body terms (up to the statistical variance). This is advantageous when the focus is on learning low-body terms. We could verify the validity of our model by
checking whether the learned Liouvillian matches the derivatives of expectation values that were not used during the learning procedure~\cite{verify}.  

\subsection{Pairwise closed system of equations}\label{subsub:closed}
Stacking the two systems Eqs.~\eqref{eq:system1}-\eqref{eq:system2}, we obtain
the combined linear system
\begin{align} \label{eq:system12}
\frac{d\mathcal{O}}{dt} 
=
MX,
\end{align}
with the vectors $\mathcal{O}=(\braket{O_{c}})$, $c=1,\dots C$, and $X=(X_\ell)$,
$\ell=1,\dots,L=51$ representing the observables derivatives in each configuration, and the unknown Liouvillian parameters on the qubit pair. 
Finally, $M$ is a $C\times L$ matrix corresponding to the trace expressions, present in Eqs.~\eqref{eq:system1}-\eqref{eq:system2}.

At this point, we are ready to estimate the Liouvillian parameters via a least-squares solution
\begin{equation}
 \hat{X} = \tilde{M}^{-1}\left( \frac{d\mathcal{O}}{dt} \right),
 \label{der_inv}
\end{equation}
with the pseudo-inverse $\tilde{M}^{- 1} = (M^{\dagger} M)^{- 1} M^{\dagger}$.
Note that when the number of configurations is sufficiently high, $M$ is of full rank (see Sec.~\ref{subsec:rank}). 
In this case, 
the least-squares solution is unique, and we measure the Liouvillian vector $X$ "unambiguously".
Repeating the procedure for all the other qubit pairs $(i,j)$, 
we obtain  the entire Liouvillian.
Also, we note that each single-body parameter $h_{i,a}, d_{i,a,i,b}$ is estimated $N-1$ times (once in each qubit pair system).
In order to extract a single value, we take the mean over all these estimates.

Thanks to this \textit{pairwise} learning, the required classical memory 
is bounded by the size of $M_{max}$, the largest possible matrix $M$ with size $(C_{\max}=360, L=51)$, and also, that is independent of $N$;  
and the number of (independent) systems to solve is $N(N-1)/2$, the number of pairs. As such, the protocol scales favourably with the system size and can be used on large-scale quantum simulators. 
In the next section we show how we can access the $\mathcal{O}(t)$ experimentally, 
and in Sec.~\ref{sec:poly_interp}, how to estimate their $t=0$ derivatives.

\section{Data sampling with randomized measurements}\label{sec:exp_friendly}

Let us now discuss how to build a vector $d\mathcal{O}/dt$ 
for all qubit pairs from an experimental system, using randomized measurements~\cite{elben_randomized_2022}.
This section describes how to build a time series $\braket{O_c(t)}$, while the next one Sec.~\ref{sec:poly_interp} discusses the details of the polynomial interpolation that gives then access to derivatives at time $t=0$.

\subsection{Measurement protocol}
\begin{figure}
    \centering
    \includegraphics[width=0.48\textwidth]{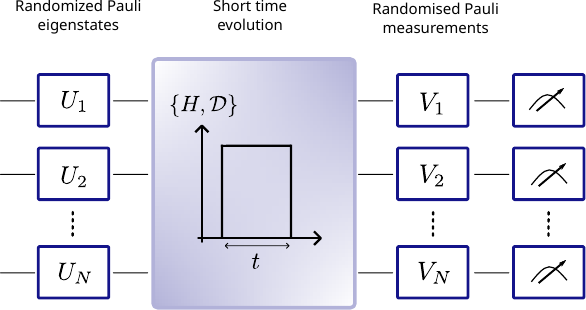}
    \caption{\textbf{Pairwise Liouvillian learning experimental protocol}. We prepare randomized Pauli states, time evolve the system and measure in random Pauli basis. We repeat the protocol for $R$ random local gates $\{ (U_i,V_i) \}$, for $N_T$ time points and $N_M$ repetitions. Then we extract the Liouvillian from the measured bitstrings.  
    }
    \label{fig:exp_protocol}
\end{figure}

Our approach is inspired from the one presented in Ref.~\cite{stilck_franca_efficient_2024}, 
and is represented schematically in Fig.~\ref{fig:exp_protocol}.
An experimental sequence begins by picking randomly two unitaries $U = \bigotimes_{k=1}^N U_k$
and $V = \bigotimes_{k=1}^N V_k$, where all single-site unitaries are sampled independently.
The unitaries $U_k$ used for state preparation are taken uniformly
in $\{ \mathcal{H}, X \mathcal{H}, \mathcal{H} S, X \mathcal{H} S, \mathbf{1}, X \}$, 
with $\mathcal{H}$ the Hadamard gate and $S$ the $\pi/2$ phase gate, to prepare the \mbox{$\pm \sigma^{(x)}$, $\pm \sigma^{(y)}$, $\pm \sigma^{(z)}$} 
Pauli eigenstates, respectively.
The unitaries $V_k$ used for the measurement step are taken in the set
$\{ \mathcal{H}, \mathcal{H} S^\dagger, \mathbf{1}\}$, in order to provide measurements
in the $\sigma^{(x)}$, $\sigma^{(y)}$, $\sigma^{(z)}$ basis respectively.

We proceed as depicted in Fig.~\ref{fig:exp_protocol}:
(i)
From the initial state $\ket{0}^{\otimes N}$, we first apply $U$.
(ii) The system is then evolved via the system's Liouvillian during a time $t$. 
(iii) Finally, we apply the unitaries $V$ and realize $N_M$ projective measurements in the computational
basis.
The sequence (i)-(ii)-(iii) is repeated for longer times $t=sdt$, $s=1,\dots,N_T$. 
We repeat this overall sequence for $R$ settings made of random choices of unitaries $(U,V)$. 
The measurement data $\mathcal{B}$ consist of a bitstring 3D array of dimensions $(N_T,R,N_M)$ 

In order to access the expectation values $\braket{O_c(t)}$ from initial mixed states
of the form Eq.~\eqref{eq:configs}, we make use of the randomization procedure. Let us consider for simplicity the preparation 
of the initial mixed state 
\mbox{$\rho_i=\tau_i\otimes (\mathbf{1}/2)^{\otimes N-1}$.} 
Let us define the subset of settings $\varrho=\bigotimes_k \tau_k$, of size $R'\le R$, such that $\tau_i=U_i\ket{0}\bra{0}U_i^\dag$.
The average quantum state prepared in the subset corresponds to the target mixed state 
\begin{equation}
    \label{eq:randominit}
\tau_i \otimes  \left(\mathbf{E}\left
[\bigotimes_{k\neq i} U_k \ket{0} \bra{0} U_k^\dag \right]
\right)
=\tau_i \otimes (\mathbf{1}/2)^{\otimes N-1}=\rho_i.
\end{equation}
We proceed similarly to reproduce the preparation of a state of the form $\rho_{ij}$.
Note that a given unitary $U$ is used multiple times in postprocessing
to effectively "prepare" various $\rho_i$, $\rho_{ij}$ over different qubits 
$i$, and qubit pairs $(i,j)$.
This parallelization is the mechanism behind the 
efficiency of randomized measurements~\cite{elben_randomized_2022} and the classical shadows~\cite{huang_predicting_2020}.

In a similar manner, we extract the expectation values $\braket{O_c}$:
among the $R'$ settings with unitaries preparing the target initial state $\rho_c$, 
we retain the $R''\le R'$ settings, 
such that the unitary $V$ provides a measurement of the operator $O_c$, the ones where $O_i = V_i^\dagger Z V_i$ (and 
\mbox{$O_j = V_j^\dagger Z V_j$}). This happens
with probability $1/3$ ($1/9$) for single-body (two-body, respectively) observables~\cite{elben_randomized_2022}.
In App.~\ref{construct_obs}, we detail the construction of such estimators for the 
measurement data $\mathcal{B}$.

\subsection{Comments about the randomization procedure}

We have described a randomized procedure to construct 
the systems Eq.~\eqref{eq:system12} for all pairs of qubits
simultaneously. In Ref.~\cite{stilck_franca_efficient_2024}, the formalism of process shadows provides an estimation for any observable $\braket{O_c}$, 
i.e. the matrix $M$ has size $C_{\max}$ and is of full rank.
In particular, in case of finite number of unitaries $R$, a configuration
$(\rho_c,O_c)$ that is incompatible with all settings $\{(U,V)\}$ will lead to the
assignment $\braket{O_c}\to 0$, even if $N_M\to \infty$. 

In our approach we only estimate the configurations $c$ that have at least one 
compatible setting $\{(U,V)\}$, ensuring that each estimation 
$\braket{O_c}$ is only affected by shot noise errors (finite value 
of $N_M$), and the randomization over initial states appearing in 
Eq.~\eqref{eq:randominit}.
As a consequence, the number of configurations $C$ becomes a random number, as does the rank of the matrix $M$.
 
In the following, we show numerically that we can obtain full-rank matrices $M$
with arbitrarily high probability and a reasonable number of settings $R$, 
thus obtaining faithful reconstructions of the Liouvillian in the large-scale scenario.

\subsection{Rank completion with finite measurement budget}\label{subsec:rank}

\begin{figure}
    \centering
    \includegraphics[scale=0.45]{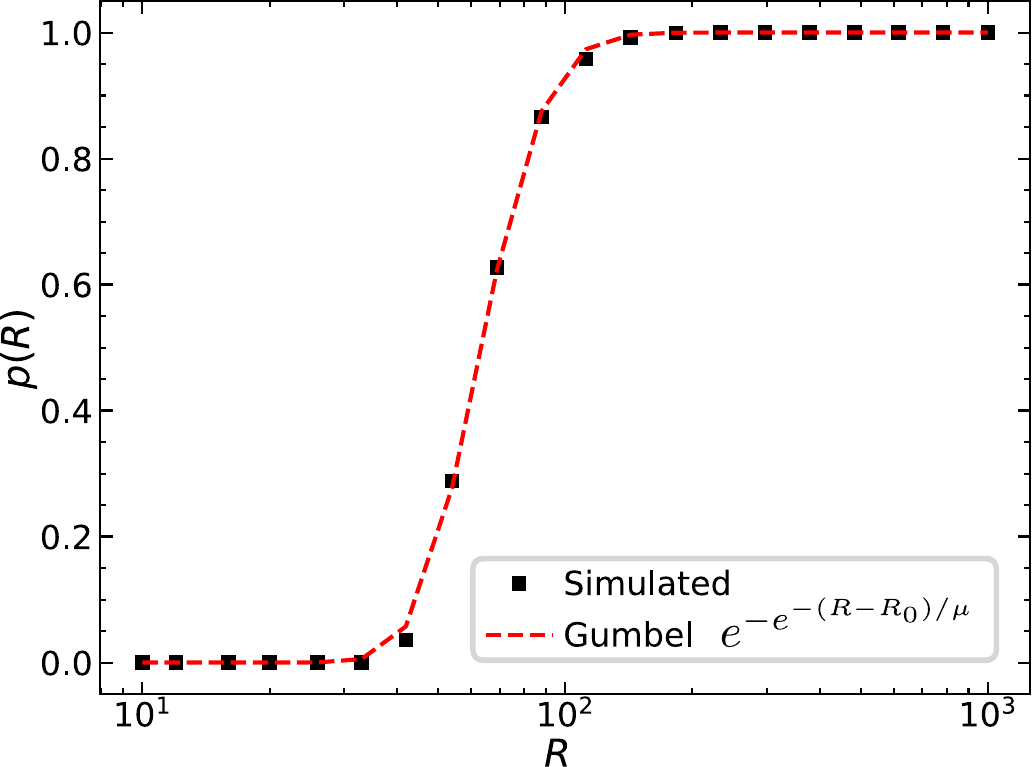}
    \caption{\textbf{Probability of having a full rank system for a given qubit pair}, as a function of the number of random Pauli unitaries, calculated numerically with $1000$ samples. We interpolated the data with a Gumbel function $e^{-e^{-(R-R_0)/\mu}}$, with $R_0=57.76$ and $\mu=15.0$. After $R>200$, the system is full rank with unity probability.}
    \label{fig:prank_2}
\end{figure}

\begin{figure}
    \centering
    \includegraphics[scale=0.6]{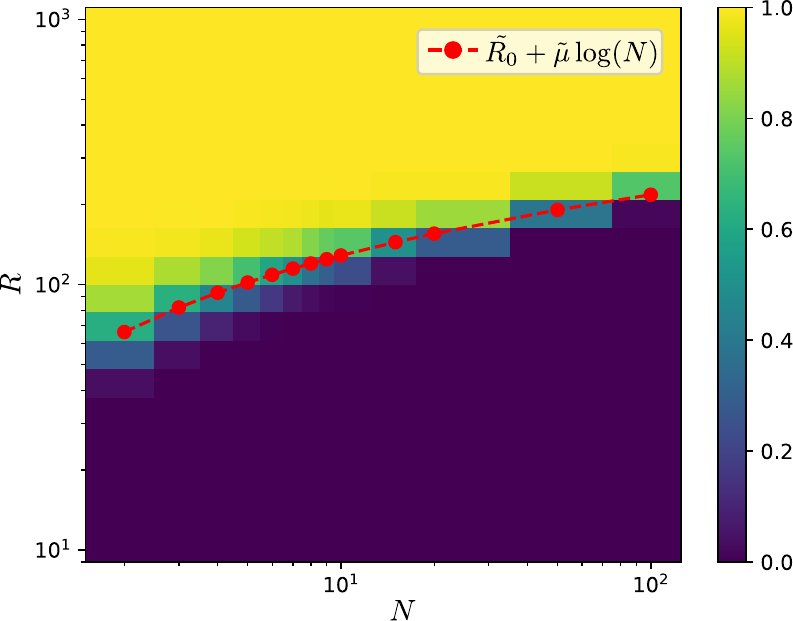}
    \caption{\textbf{Probability of having a full rank system on all the qubit pairs}, as a function of the number of random Pauli unitaries, calculated numerically with $1000$ samples. We interpolated the isoprobability curve $p(R,N)=0.5$, and obtained a logarithmic dependence $R = \tilde{R}_0 + \tilde{\mu} \text{ log}(N)$ of the parameters, with $\tilde{R}_0, \tilde{\mu} = 39.31, 38.84$. This illustrates the scalability of the full-rank condition with the system size.}
    \label{fig:prank_N}
\end{figure}
We begin our numerical analysis by studying the probability $p(R)$ that the matrix $M$
is full rank for a given qubit pair $(i,j)$, as a function of the number 
of settings $R$.
The results shown in Fig.~\ref{fig:prank_2} reveal that the probability resembles a threshold 
activation function. For $R<29$, we have $p(R)=0$, meaning that for such settings 
there exist no sets of configurations
with full rank. 
On the other hand, for $R>150$, $p(R) > 0.99$, meaning that almost all 
such settings provide full-rank matrices $M$. 
Overall, we observe that $p(R)$ can be well fitted with a Gumbel function 
$f(R)=\exp(-\exp(-(R-R_0)/\mu))$, even though we do not have strictly $f(R<29)=0$.

Moving to the multi-pair case,
we show in Fig.~\ref{fig:prank_N}
the probability $p(R,N)$ that the $R$ settings of a $N$-qubit system provide 
full-rank matrices $M$ for all $N(N-1)/2$ qubit pairs.
For each value of $N$, we observe the same kind of threshold activation as seen in 
Fig.~\ref{fig:prank_2}.
Moreover, the location of the threshold scales approximately logarithmically with $N$. 
This highlights a crucial scalable aspect of the randomization procedure:  for $R=\mathcal{O}(\log(N))$, we can ensure with overwhelming probability a  complete, pairwise, full-rank, Liouvillian learning of our qubit system. 

This result can be understood as follows.
In full rigor, the rank of two matrices $M$, $M'$, associated with two qubit pairs $(i,j)$, and $(i',j')$ respectively may be statistically correlated when the pairs are not fully distinct.
However, neglecting this effect and describing all full-rank conditions as independent processes, we obtain $p(R,N)=p(R)^{N(N-1)/2}$.
Using $p(R)\approx f(R)$, we obtain for the isoprobability $p(R, N) = \delta \in \ ]0,1[$,
\begin{equation}
R = R_0 +  \mu \big( \log(\frac{N(N-1)}{2}) - \log(|\log(\delta)|) \big) =\mathcal{O}(\log(N)).
\end{equation}
We note that the threshold at $R=29$ reveals the existence of
a minimal set of configurations required to satisfy the full-rank condition (on a single qubit pair).
For future work, it would be interesting to investigate whether this set of configurations
can be obtained deterministically and simultaneously for all qubit pairs,
in connection with the concept of overlapping state tomography~\cite{cotler_quantum_2020}.
Finally, the full-rank condition can be achieved faster~\cite{verify} when considering a reduced Liouvillian ansatz, since the coefficients $R_0$ and $\mu$ depend only on the ansatz and the set of configurations $(\rho_c, O_c)$.

\section{Polynomial interpolation for parameter extraction}\label{sec:poly_interp}
While the system Eq.~\eqref{eq:system12} is expressed in terms 
of derivatives $d\mathcal{O}/dt = \{ d\braket{O_c}/dt \}$ at time $t=0$, we have described a measurement and postprocessing procedure giving access to the time series $\{\braket{O_c(t)}\}$, for 
$t=sdt$, $s=1,\dots, N_T$.
In the following, we describe how to fill this gap via polynomial interpolation, and discuss in particular the respective roles
of systematic (fitting) and statistical errors in the overall procedure.

\subsection{Defining the observables to be interpolated}
First, to facilitate the interpolation, we find it convenient to recast the least-squares solution $\hat{X}$ given by Eq.~\eqref{der_inv} by applying $\tilde{M}^{-1}$ before the derivation operation,
\begin{equation}
	 \hat{X} =  \frac{d}{dt}
	 \left(\tilde{M}^{-1}  \mathcal{O}(t)\right) = \frac{dY(t)}{dt},
	 \label{poly_ols}
\end{equation}
with 
\begin{equation}
	Y(t) = \tilde{M}^{-1}  \mathcal{O}(t),
	\label{syst_ij_poly}
\end{equation}
which is a vector of $L$ time-dependent functions. 

\begin{figure}
    \centering
    \includegraphics[width=0.48\textwidth]{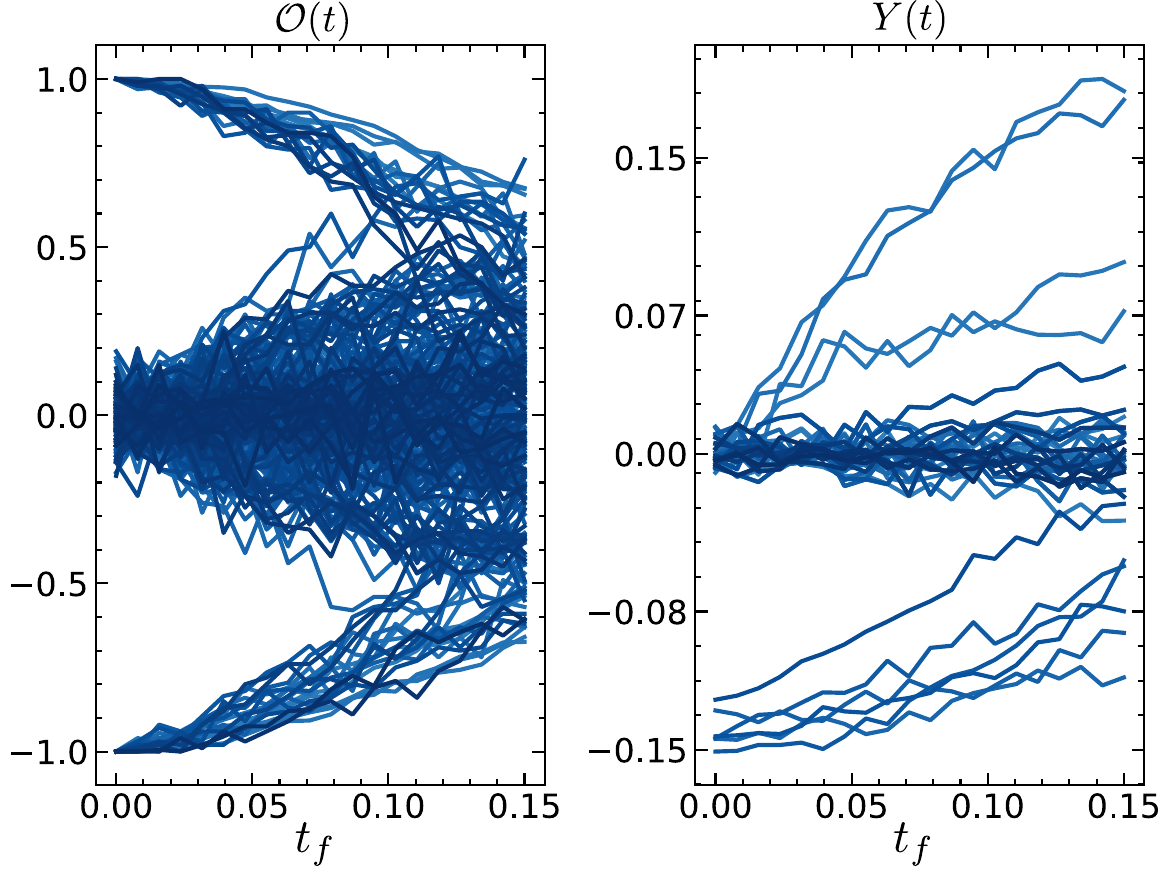   } 
    \caption{\textbf{Time dynamics of $\mathcal{O}(t)$ (left) and $Y(t)$ (right).} 
	The model and parameters are given in Sec.~\ref{sec:numerical_study}.
	Applying the inversion matrix $\tilde{M}^{-1}$ 
	to the measured observables $\mathcal{O}(t)$, we obtain a few curves $Y(t)$ with reduced statistical noise and well-suited for polynomial interpolation. 
    }
    \label{obs_vs_polys}
\end{figure}

From Sec.~\ref{sec:exp_friendly}, we have from an experimental dataset $\mathcal{B}$ access to the time series
$\{Y(t)\}=\{\tilde{M}^{-1}  \mathcal{O}(t) \}$, for $t=sdt$, $s=1,\dots, N_T$, with maximal time $t_f = N_T dt$.
From polynomial interpolation, we can then approximate the continuous function $Y(t)$, and its derivative $dY(t)/dt$ at time $t=0$. 
Compared to the polynomial interpolation of the vector $\mathcal{O}$, this approach is advantageous because we have $L=51$ time series instead of $C\lesssim 360$ to analyze. In particular $Y_\ell$, an element of $Y$, built as a linear combination over the vector $\mathcal{O}$, has typically a smaller statistical error than $\braket{O_c}$, an element of $\mathcal{O}$, which will facilitate the polynomial interpolation procedure described below. 
In Fig.~\ref{obs_vs_polys} we show examples of time series $\mathcal{O}(t)$ (left) and $Y(t)$ (right), with the model and parameters specified in Sec.~\ref{section:code}. While the $C$ curves $\mathcal{O}(t)$ do not provide direct insight about the operator content of the Liouvillian, the derivatives at time $t=0$ of $Y(t)$ will each correspond to a term in the Liouvillian and we can visually inspect to what extent each curve $Y(t)$ can be fitted via a low degree polynomial.

\subsection{Assessing polynomial interpolation}

For each value of $\ell=1,\dots,L$, the time series $Y_\ell(dt), \dots, Y_\ell(t_f)$   is thus 
fitted with a polynomial of degree $D_{\ell}$ that we write
\begin{equation}
    q(t) = \sum_{d=0}^{D_\ell} \beta_d t^d,
\end{equation}
and use $\beta_1$ as estimate for $dY_\ell(t)/dt$ at $t=0$.
The least-squares coefficients of such polynomial are expressed as
\begin{equation}
    \beta = \tilde{V}^{-1}Y_\ell(t),
\end{equation}
with $\beta = (\beta_d)$, and $\tilde{V}^{-1}$ a matrix of dimension \mbox{$(D_\ell+1,N_T)$}, which is the pseudo-inverse of the Vandermonde matrix
\mbox{$[V]_{s,d}=(dt s)^d$}, 
\mbox{$s=1,\dots,N_T$}, \mbox{$d=0,\dots,D_{\ell}$}.

The choice of the degree $D_\ell$ used to interpolate a given time series $Y_\ell(t)$ results from a tradeoff in dealing with systematic (fitting) errors and statistical errors, which is a well-known effect in least-squares  regression~\cite{hastie_stat_learning_2009}. 
At low degrees $D_\ell$, fitting errors dominate. 
On the other hand, as $D_\ell$ increases, the propagation of statistical errors
from each time point $Y_\ell(t)$ to $\beta_1$ is more and 
more amplified. This effect will be commented in detail in the next section that is devoted to numerical simulations of the overall Liouvillian procedure.

In order to automate the procedure for degree selection, we use $K$-fold cross-validation~\cite{hastie_stat_learning_2009}. 
The dataset (time points) is randomly partitioned into $K$ equally sized subsets (folds).
We interpolate the $Y_\ell(t)$ on $K-1$ folds, extrapolate the obtained polynomial on the remaining fold, and compute the residuals with respect to $Y_\ell(t)$ on this fold. We repeat this procedure $K$ times and pick the degree that has the smallest residuals on average.

\section{Numerical tests of pariwise Liouvillian learning}\label{sec:numerical_study}
Our numerical study aims at showing the performance 
of our Liouvillian learning protocol in experimentally realistic situations.
In particular, we would like to provide guidance on choosing the best parameters to maintain both systematic and statistical errors below a certain level.

To specifically target trapped ions experiments~\cite{kraft_bounded-error_2025}, we consider an $XY$ model of size $N=2,\dots,10$ with power-law interactions ($\hbar=1$) 
\begin{equation}
H = \frac{J}{2}
\left(\sum_{i<j}\frac{1}{|i-j|^\alpha}	(\sigma_i^{(x)} \sigma_j^{(x)} + \sigma_i^{(y)} \sigma_j^{(y)})\right) + B \sum_i \sigma_i^{(z)},
\label{eq_h_ions}
\end{equation}
where $J,B = 4,1 $ are fixed in our simulations, setting the time unit, and with $\alpha=1.5$.  
We will show simulation results including dissipation in Sec.~\ref{section:code}.

In our simulations, we used $R=1000$ random unitaries, $N_M=1000$ projective measurements per unitary, and $N_T=40$ time points. 
To provide an estimation of statistical error bars, each simulation was repeated $20$ times. We describe in Sec.~\ref{section:code} how we carried out the simulations.

\subsection{Analysis for a single pair}
\begin{figure}
	\includegraphics[width=0.45\textwidth]{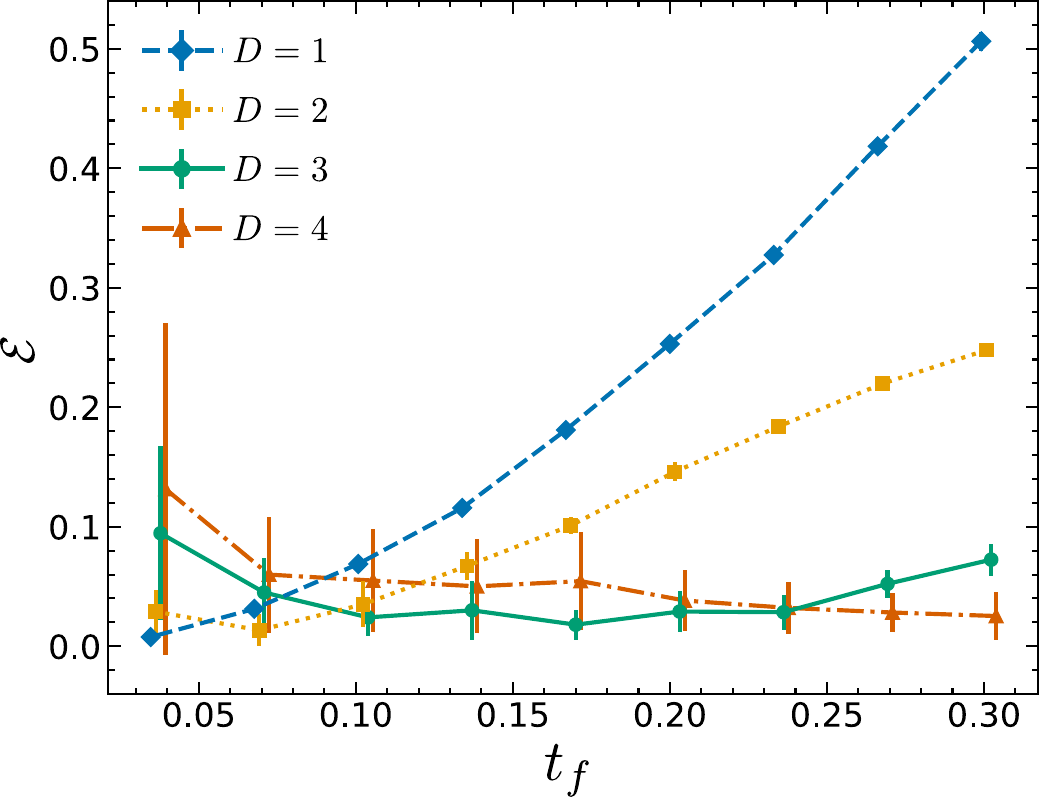}
	\caption{\textbf{Reconstruction error as a function of the maximal time and interpolation degree.} We simulated numerically the protocol with the parameters $N=2$, \mbox{$R=1000$}, $N_M = 1000$, $N_T = 40$, with a trapped ion Hamiltonian, and averaged the data over $20$ samples. This illustrates a bias (low degree, long times) variance (high degree, short times) tradeoff.
	}
	\label{fig:time_res}
\end{figure}

In Fig.~\ref{fig:time_res}, we study the
reconstruction error \mbox{$\mathcal{E}=\mathbf{E}\left[||\hat{X}-X||_1\right]$}, where $\mathbf{E}$ is the ensemble average. We consider a single pair $N=2$ for various evolution time $t_f$. The degrees $D=1,2,3,4$ of interpolation were taken to be the same for all entries $Y_\ell$.
For $D=1,2$ (blue and yellow curves),
we see that the error is at first close to zero, then increases due to fitting errors.
Higher degrees $3$ and $4$ (green and red) are less prone to fitting errors, but suffer more from statistical errors at short evolution time $J dt\ll 1$, due to the small value of the resolution time, the Vandermonde matrix becomes ill-conditioned~\cite{walter_vandermonde_2020}.

\subsection{Analysis for the full system}
\begin{figure}
	\includegraphics[width=0.43 \textwidth]{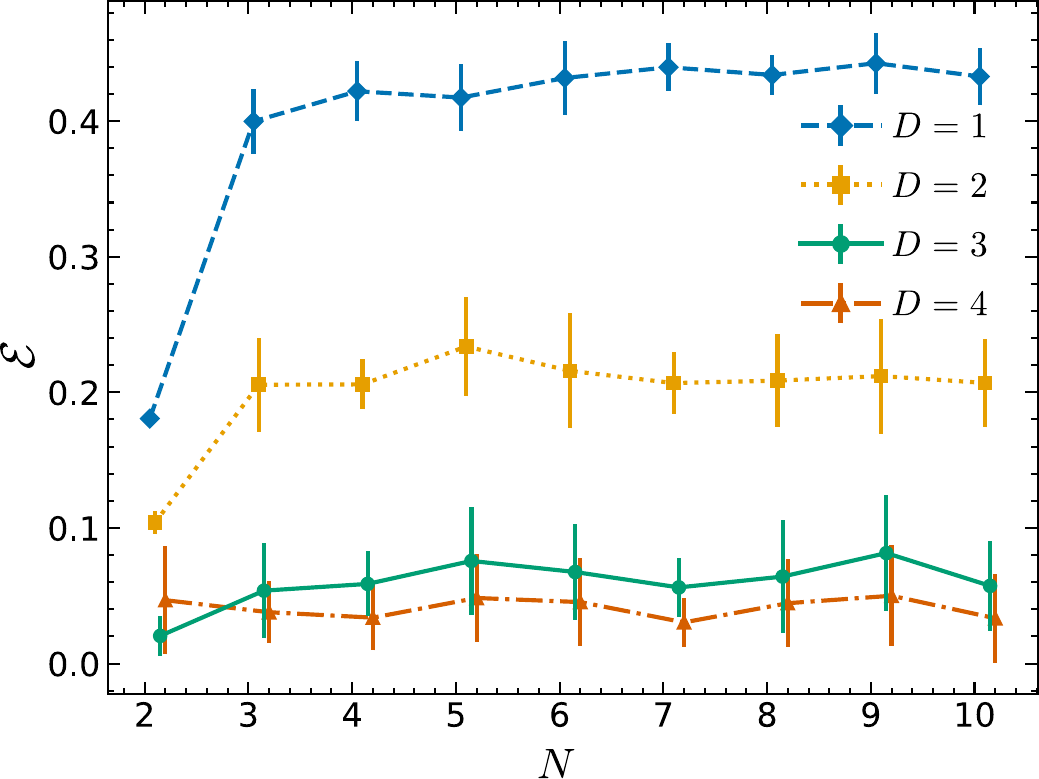}
	\caption{\textbf{Reconstruction error as a function of the system size and interpolation degree.} We simulated numerically the protocol with the parameters $R=1000$, \mbox{$N_M = 1000$}, $N_T = 40$, $t_f = 0.15$, with a trapped ion Hamiltonian with power-law decaying interactions with $\alpha = 1.5$, and averaged the data over $20$ samples. 
	This shows the saturation of the errors with the system size.
	}
	\label{fig:err_deg_N}
\end{figure}

In Fig.~\ref{fig:err_deg_N}, we study the reconstruction error $\mathcal{E}$, averaged over all qubit pairs, as a function of the system size $N$ and degree $D$. 
Looking at the degree $1$ (blue), we see that the error jumps between $N=2$ to $3$, then increases slowly for $N \geq 3$. Degree $2$ and $3$ (yellow and green) also show a jump between $N=2$ to $3$, but then stay still (up to the error bars), and degree $4$ stays approximately constant.
This saturation puts into evidence the scalable reconstruction of the overall procedure. 
For a given measurement budget, the average reconstruction error does not scale at all with system size, asymptotically.

We attribute the origins of the jumps to the randomization procedure of initial states via the preparation unitaries $U$, as shown in Eq.~\eqref{eq:randominit}.
Let us consider Liouvillian learning on the first pair of qubits $(1,2)$ for simplicity. 
For $N=2$, there is no statistical error in the randomization in  Eq.~\eqref{eq:randominit}.
For $N=3$, the state of one  qubit $j=3$ must be randomized to effectively prepare the mixed state $\mathbf{1}/2$. The jump from $N=2$ to $N=3$ corresponds thus to a statistical error in the preparation of the "spectator" qubit $j=3$. 
For $N>3$, we have more spectator qubits, but the overall effect of an imperfect randomization saturates. 
Intuitively, this is because, as interactions decay with distance, the state of the spectator qubits $j$ that are far away from the pair $(1,2)$ can only marginally affect the short-time dynamics $\braket{O_c(t)}$. 
Note that Ref.~\cite{stilck_franca_efficient_2024} has rigorously established the scalability of the polynomial interpolation, based on such a notion of local interactions, and in particular on the existence of Lieb-Robinson bounds.
   
The fact that the reconstruction error is asymptotically independent of the qubit number 
has also practical consequences for choosing the crucial parameters $R,N_M,N_T,t_f,D$. One can benchmark a reconstruction with $N=2-5$, and choose optimal parameters that will remain meaningful for any system size.

\section{Codes for pairwise Liouvillian learning}\label{section:code}

In this last section, we briefly explain how our codes, available on our GitLab repository~\cite{WLAM_git}, work, 
and how to use them to analyze data from quantum simulation experiments. 

\begin{figure}
    \centering
    	\includegraphics[width=0.50\textwidth]{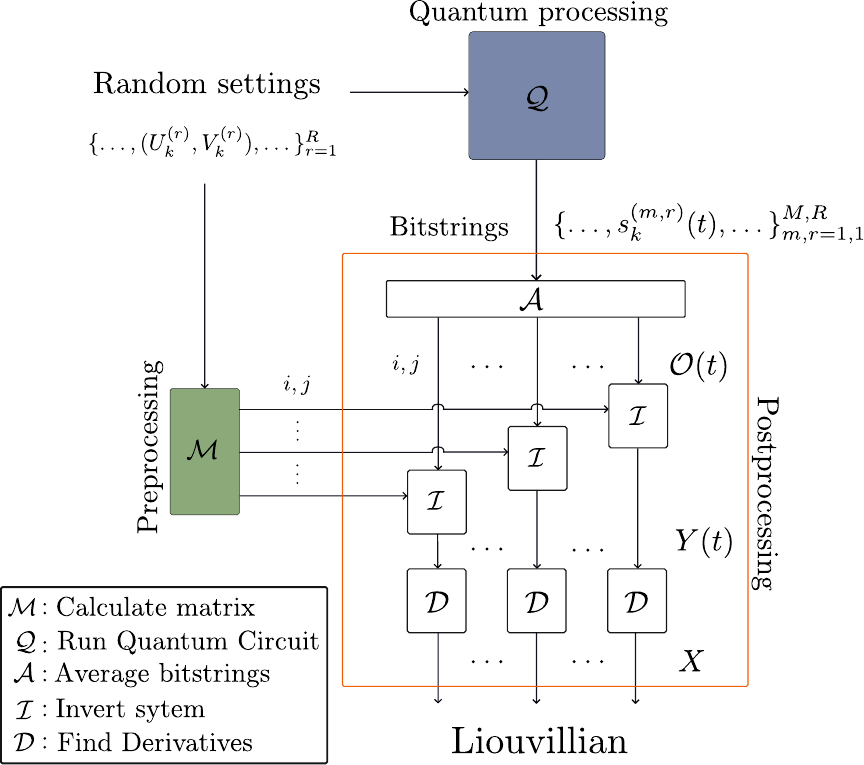}
    \caption{Schematic diagram of the Liouvillian learning protocol: 
    we draw $R$ random settings; 
    we compute pairwise matrices in a preprocessing step; 
    we run the quantum processing with $M$ repetitions and obtain bitstrings; 
    we post-process the data to extract the Liouvillian coefficients.
	}
    \label{HLL_diagram}
\end{figure}

The main steps are illustrated in Fig.~\ref{HLL_diagram}. 
First, we pick a sequence of $R$ random Pauli states and measurement basis 
\mbox{$\{ \dots,(U_k^{(r)}, V_k^{(r)}),\dots \}_{r=1}^R$}, 
from which we generate \texttt{Qiskit.QuantumCircuit} instructions that need to be applied before and after the Liouvillian time evolution, as depicted in Fig.~\ref{fig:exp_protocol}.

Then we can run the experiments in the quantum simulator, or simulate them classically, for the $R$ settings, the $N_T$ (short) time evolution and $N_M$ repetitions, and obtain the bitstrings $\{ \dots, s_k^{(m,r)}(t),\dots \}_{m,r=1,1}^{M,R}$, which will serve as input in the post-processing step.
In our repository, we also provide code to simulate any two-body Liouvillian dynamics (with local noise), via \texttt{Qiskit} and using a trotterized approximation. Although we focused on AQS in the introduction, we provide here quantum circuits, in case one wants to run the protocol in a digital quantum computer setting. 

Then, for each qubit pair, 
we compute and average the one- and two-body time sequences $\mathcal{O}(t)$ from the bitstrings. The corresponding pairwise matrix $M$ is efficiently constructed by slicing the matrix $M_{max}$, that we have already precomputed. 
We invert the system to obtain the $Y$, that we interpolate to estimate the derivatives, and so get the $X$.
The degree of interpolation is selected from the list of degrees, typically  \texttt{Deg = 1-5}, and using $K-$fold cross-validation, with $K=3$.

\begin{figure}
    \vspace*{2.5ex}  
    	\includegraphics[width=0.50\textwidth]{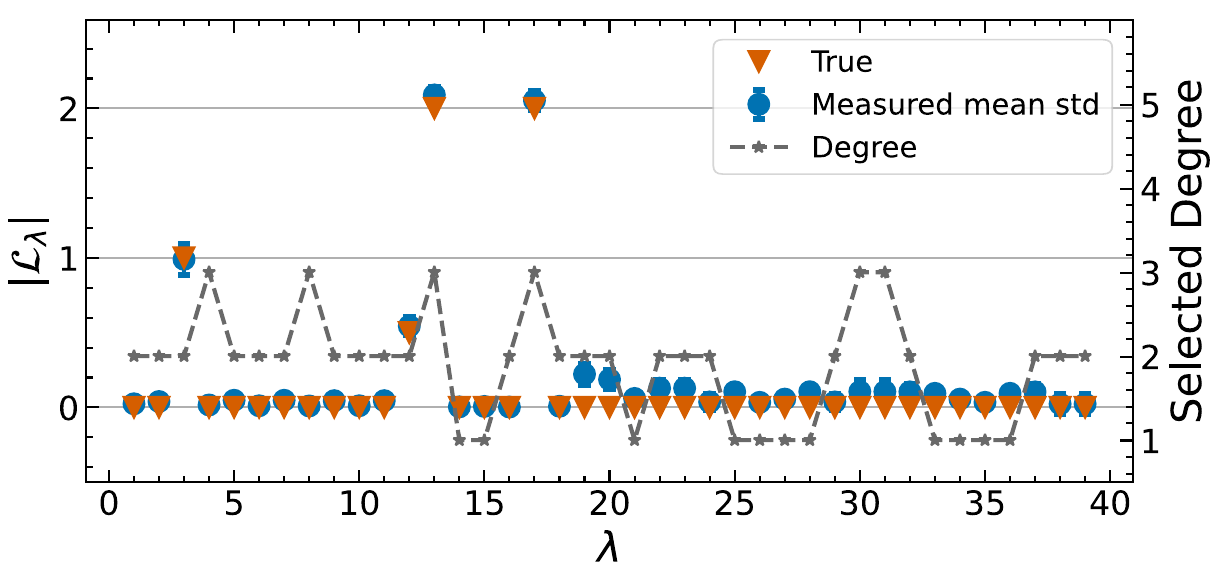}
    \caption{\textbf{Absolute value of the learned Liouvillian coefficients with the interpolation degree}, averaged over the qubits and nearest-neighbor qubits. We show in blue the estimated data and in orange the true values. This illustrates that the protocol can correctly learn the Liouvillian.}
    \label{fig:Learned_terms}
\end{figure}

We implemented in Fig.~\ref{fig:Learned_terms} the evolution of Eq.~\eqref{eq_h_ions}, with $J/2=2, \ B=1$, and, here, with a local dephasing noise model with $\gamma = d_{i,z,i,z}=0.5$ for $i =1,\dots, N$. We used a reduced experimental budget compared to the numerical study in Sec.~\ref{sec:numerical_study}, with $N=10$, \mbox{$R=800$}, $N_M=200$, $N_T=40$, $t_f=0.1$. This result in a total of $6.4$ millions measurements; but $t_f$ being small, it only represents $3.2$ hundred thousand quantum evolution time (in the time unit of the interactions $J$ and $B$). 

After filling the file \texttt{task.json}, to specify all the (above) simulation and protocol parameters, we ran the commands
\begin{lstlisting}[language=Python,backgroundcolor=\color{backcolor},basicstyle=\ttfamily\footnotesize]
python main_simu.py task.json
python main_pairwise_learning.py task.json
\end{lstlisting}
to simulate and learn the Liouvillian coefficients.
To analyze the learned data, and in particular, reproduce Fig.~\ref{obs_vs_polys} and Fig.~\ref{fig:Learned_terms},  we used the notebook \texttt{Liouv\_learning.ipynb}, that is also available in the repository.

Having determined $\mathcal{O}(N^2)$ coefficients, $1335$ here, we only show, for simplicity, the averaged one-body terms, and averaged two-body terms over nearest-neighbor pairs, and their error bars. 
The indices 
$\lambda \in [1-3], [4-12], [13-21], [22-39]$ describe the 
$\mathcal{L}_\lambda \in \{\overline{h}_{i,a}\}_a, \{\overline{d}_{i,a,i,b}\}_{a,b}, \{\overline{h}_{i,a,j,b}\}_{a,b}, \{\overline{d}_{i,a,j,b}\}_{a,b}$ coefficients, respectively.
The dominant contributions at $\lambda = 3, 12, 13, 17$, corresponding to 
$B$, $\gamma$, $h_{i,x,j,x}$, $h_{i,y,j,y}$ 
are correctly learned, with values compatible with the expected ones within error bars; while the remaining terms are small. 
Furthermore, we analyzed the decay of the coefficients $h_{i,x,j,x}, h_{i,y,j,y}$ as a function of the distance between the qubits, and found approximately the correct powerlaw parameters $J=2.08 \pm 0.04, \ \alpha=1.51 \pm 0.06$, illustrating the protocol's capability to learn long-range couplings. 

To finish, we note that Ref.~\cite{kraft_bounded-error_2025} shows how to
analyze the spatial structure of both Hamiltonian and
dissipation terms, and to apply regularization techniques
to force a particular profile in the coefficients.

\section{Conclusion and Outlook}

We have studied in detail the pairwise Liouvillian learning protocol from randomized measurements. 
We have reviewed each step of data acquisition and postprocessing and shown with numerical simulations the scalability aspect of the overall procedure. 
In particular, we have shown that the required classical memory is independent of the system size, and we have quantified the influence of the key parameters on the total reconstruction error. This provides useful guidelines for experimentalists aiming to implement the protocol. 

The simplicity and the overall performance of our protocol
make it possible to consider several extensions.
First of all, as randomized measurements are informationally complete, the assumption of a certain ansatz Eqs.~\eqref{H_model}-\eqref{D_model} can in principle be relaxed. It would thus be interesting to formulate 
ansatz-free Liouvillian learning in this context, as was done for Hamiltonian learning in Ref.~\cite{hu_ansatz-free_2025}, and study the practical aspects of the protocol. Exploiting the time series $\mathcal{O}(t)$ beyond derivatives at time $t=0$, we would also like to predict the experimental effort that would be needed to reconstruct faithfully a time-dependent Liouvillian~\cite{franca_learning_2025} from randomized measurements. 
Finally, we could explore methods to reach Heisenberg-limited scaling, which is the optimal metrology precision, see Ref.~\cite{hu_ansatz-free_2025, brahmachari_learning_2026}. 

\section*{Acknowledgments}
We thank P. Zoller for feedbacks on the manuscript,
and T. Kraft, B. Kraus, C. Roos, M. Filippone, R. Whitney for useful discussions.
We acknowledge funding from the French
“Plan France 2030” research programme HQI
(Grant No. ANR-22-PNCQ-0002).
W.L. is also supported by the 
“QuanTEdu-France” programme 
(Grant No. ANR-22-CMAS-0001, France 2030).
D.S.F. acknowledges financial support from the Novo Nordisk Foundation (Grant No. NNF20OC0059939 Quantum for Life). D.S.F. and Z.S. acknowledge support by the ERC grant GIFNEQ 101163938.


%

\appendix
\section{Estimation of the observables from randomized measurements}\label{construct_obs}

This appendix provides estimation formulas for the observables $\braket{\mathcal{O}(t)}$
based on the measured bitstrings of the randomized procedure described in the main text.

We start by describing the acquisition of local operators from $N$-qubit bitstrings, then describe the preparation of mixed states, and finally merge the two methods to obtain the final estimator.

\subsubsection{Estimating local operators}
We consider a setting where we prepare a fixed state $\rho(t)$, apply $V=\bigotimes_{k=1}^N V_k$ on each qubit, and measure in the $Z$ basis. 
We repeat $N_M$ times the measurement and obtain the corresponding bitstrings $\{ \mathbf{s}^{(m)}(t) \}_{m=1}^{N_M}$, with $\mathbf{s}^{(m)}(t)=(s_1^{(m)}(t),\dots,s_N^{(m)}(t))\in \{0,1\}^{\otimes N}$.
Let's consider first a single-body operator $O=\sigma_i$, with 
\mbox{$\sigma_i=W^\dagger_i Z_i W_i$} a Pauli operator on site $k=i$. From the measured data, we want to estimate tr$(\rho(t) O)$. Let us denote $\braket{Z_i(t)}$ the expectation value of $Z_i$ on the rotated state $V\rho(t) V^\dag$ that we measured. For the case of a setting $V$ that is compatible with the measurement of $\sigma_i$, in the sense that $V_i=W_i$, we find that,
\begin{equation}
\begin{split}
\braket{Z_i(t)}&=\mathrm{tr}(V \rho(t) V^\dagger \ Z_i ) \\
&=\mathrm{tr}(\rho(t) \ V^\dagger_i Z_i V_i 
\bigotimes_{k \neq i} V^\dagger_k V_k) \\
&=\mathrm{tr}(\rho(t) \ 
\sigma_i \otimes (\mathbf{1}/2)^{\otimes N-1})  \\
&=\mathrm{tr}(\rho(t) O).
\end{split}
\end{equation}
On the other hand, we estimate the left-hand side from the bitstrings using Born probabilities, we have that
\begin{equation}
P(\mathbf{s}|V,\rho(t))=
\frac{1}{M}\sum_{\mathbf{s}^{(m)}(t)}
\delta_{\mathbf{s},\mathbf{s}^{(m)}(t)},
\end{equation}
where $\mathbf{s}=(s_1,...,s_N)$, represents all the possible bitstrings. Thus, if $V_i=W_i$, we define $\braket{\hat{Z_i}(t)}$ as the estimator of $\mathrm{tr}(\rho(t) O)$, given by
\begin{equation}
\braket{\hat{Z_i}(t)}
=\sum_{\mathbf{s}} (-1)^{s_i} P(\mathbf{s}|V,\rho(t)).
\end{equation}
Similarly, if we want to evaluate two-body terms of the form $O=W_i^\dagger Z_i W_i \ W_j^\dagger Z_j W_j$, the compatible measurements are those satisfying the conditions $V_i=W_i$ and $V_j=W_j$. We define $\braket{\hat{Z_{ij}}}(t)$ as the estimator of $\mathrm{tr}(\rho(t) O)$, given by
\begin{equation}
 \braket{\hat{Z_{ij}}(t)}
=\sum_{\mathbf{s}} (-1)^{s_i+s_j} P(\mathbf{s}|V,\rho(t)).
\end{equation}

\subsubsection{Measuring expectation values from initially mixed states}
Let us consider first a single-body Pauli state 
\mbox{$\rho=\tau_i  \otimes (\mathbf{1}/2)^{\otimes N-1}$}, 
with $\tau_i=G_i \ket{0} \bra{0} G_i^\dag$,
and let us fix the basis of measurement $V$. 
Because $\rho$ involves mixed states $\mathbf{1}/2$, 
the expectation value 
\mbox{$\braket{Z_i(t)}=\mathrm{tr}(\rho(t) V_i^\dagger Z_i V_i)$} 
is not directly measurable in a quantum simulation experiment. 
Instead, we prepare many random pure states of the form 
\mbox{$\varrho=\bigotimes_k \varrho_k$}, with 
\mbox{$\varrho_k = U_k \ket{0} \bra{0} U_k^\dag$} an eigenstate of a Pauli operator. 
The $U_k$ are drawn randomly in 
\mbox{$\{ \mathcal{H}, X \mathcal{H}, \mathcal{H} S, X \mathcal{H} S, \mathbf{1}, X \}$}. 
Let's denote \mbox{$\braket{Z_i(t)}_{\varrho}=\mathrm{tr}(\varrho(t) V_i^\dagger Z_iV_i)$} 
the expectation value of $Z_i$ with the initial state $\varrho$, 
and consider a setting where $\varrho$ is compatible with the preparation of $\rho$, 
in the sense that $U_i=G_i$. We find
\begin{equation}
\begin{split}
    \mathbf{E}
    \left[
    \braket{Z_i(t)}_{\varrho}
    \right]=&
    \mathrm{tr}
    \left(
    e^{\mathcal{L}t}
    \mathbf{E}[\varrho] V_i^\dagger Z_i V_i
    \right)\\
    =&
    \mathrm{tr}
    \left(
    e^{\mathcal{L}t}
    \tau_i \bigotimes_k \mathbf{E} \left[ \varrho_k \right] V_i^\dag Z_i V_i
    \right) \\
    =& \braket{Z_i(t)},
\end{split}
\end{equation}
where $\mathcal{L}$ is the Liouvillian, and $\mathbf{E}$ denotes the average over compatible $\varrho$. 

This can be shown using the equivalent writing \mbox{$\varrho_k=U_k \ket{0} \bra{0} U_k^\dag = \frac{\mathbf{1}+\epsilon_k b_k}{2}$} 
with $\epsilon_k$ in $\{ \pm1 \}$ and the $b_k$ in $\{\sigma_k^{(x)},\sigma_k^{(y)},\sigma_k^{(z)}\}$. This results in 
\begin{equation}
\begin{split}
\mathbf{E}[\varrho_k]
=&\mathbf{E}\left[\frac{\mathbf{1}+\epsilon_k b_k}{2}\right] \\
=&\frac{1}{6}
\sum_{b_k}\sum_{\epsilon_k=\pm1}
\frac{\mathbf{1}+\epsilon_k b_k}{2}=\mathbf{1}/2.
\end{split}
\end{equation}
Now, to evaluate two-body expectation values with initial state $\rho=\tau_i \otimes \tau_j  \otimes (\mathbf{1}/2)^{\otimes N-1}$, we use the compatible state-preparation unitaries that satisfy $U_i=G_i$ and \mbox{$U_j=G_j$}. The average over such $U$ gives
\begin{equation}
    \mathbf{E}
    \left[
    \braket{Z_{ij}(t)}_{\varrho}
    \right]
    =\braket{Z_{ij}(t)}.
\end{equation}

\subsubsection{Randomized states and measurements}\label{RSM}
Now we want to estimate $\mathrm{tr}(\rho(t) O)$, 
with the single-body setting 
$\rho = G_i \ket{0} \bra{0} G_i^\dag \otimes (\mathbf{1}/2)^{\otimes N-1}$ 
and \mbox{$O=W^\dagger_i Z_i W_i$}, that we label as $c_i=(G_i, W_i)$. 
Following Ref.~\citep{stilck_franca_efficient_2024}, 
we prepare $R$ random experimental settings 
$\{ (U^{(r)},V^{(r)}) \}_{r=1}^{R}$, 
with $U^{(r)}=\bigotimes_{k=1}^N U_k^{(r)}$, 
and \mbox{$V^{(r)}=\bigotimes_{k=1}^N V_k^{(r)}$}. 
We repeat the experiment (initialisation, evolution, measurement) $N_M$ times for each setting and obtain the corresponding bitstrings 
$\{ \mathbf{s}^{(m,r)}(t) \}_{m,r}$, 
with $\mathbf{s}^{(m,r)}(t)=(s_1^{(m,r)}(t),\dots,s_N^{(m,r)}(t))$. 
We obtain
\begin{equation}
\braket{\hat{Z_i}(t)}^{(r)}
=\sum_{\mathbf{s}} (-1)^{s_i} P(\mathbf{s}|r,t),
\end{equation}
with 
\mbox{$
P(\mathbf{s}|r,t)=
\frac{1}{M}\sum_{\mathbf{s}^{(m,r)}(t)}
\delta_{\mathbf{s},\mathbf{s}^{(m, r)}(t)},
$} 
and \mbox{$\mathbf{s}=s_1,...,s_N$}.
By combining the conditions of the two previous sections, on the compatible states and the compatible unitaries, we can average over the random configurations and build the estimator $\hat{O}_{c_i}(t)$ of $\mathrm{tr}(\rho(t) O)$ such as:\\
if $\sum_r \delta^{(c_i, r)} \geq 1$,
\begin{equation}
\hat{O}_{c_i}(t)=\frac{\sum_r \delta^{(c_i, r)}  \braket{\hat{Z_i}(t)}^{(r)}}{\sum_r \delta^{(c_i, r)}},
\end{equation}
with $\delta^{(c_i, r)} = \delta_{G_i, U_i^{(r)}} \delta_{W_i, V_i^{(r)}}$. 

For a two-body setting $c_{ij}=(G_i, W_i, G_j, W_j)$, we define
\begin{equation}
\braket{\hat{Z_{ij}}(t)}^{(r)}
=\sum_{\mathbf{s}} (-1)^{s_i+s_j} P(\mathbf{s}|r,t),
\end{equation}
and use the estimator $\hat{O}_{c_{ij}}(t)$ such as:\\
if $\sum_r \delta^{(c_{ij}, r)} \geq 1$,
\begin{equation}
\hat{O}_{c_{ij}}(t)=\frac{\sum_r \delta^{(c_{ij}, r)}  \braket{\hat{Z_{ij}}(t)}^{(r)}}{\sum_r \delta^{(c_{ij}, r)}},
\end{equation}
with \\
$\delta^{(c_{ij}, r)}=\delta_{G_i, U_i^{(r)}} \delta_{W_i, V_i^{(r)}} \delta_{G_j, U_j^{(r)}} \delta_{W_j, V_j^{(r)}}$.

Now gathering all settings $c = c_i, c_j, c_{ij}$, 
that were 'observed', 
in the sense that $\sum_r \delta^{(c, r)} \geq 1$ (we discard the ones where $\sum_r \delta^{(c, r)} = 0$),
we construct $\mathcal{O}(t)$, and call $C \leq R$ the number of settings for this qubit pair.

\end{document}